\documentclass[aip, 11pt,graphicx]{revtex4-2}

\usepackage{graphicx}
\usepackage{amsmath}
\usepackage{bm}
\usepackage{dcolumn}
\usepackage{bm}
\usepackage{xcolor, soul}
\usepackage{amsfonts}
\usepackage{booktabs}
\usepackage{siunitx}
\usepackage{longtable}
\usepackage[colorlinks,citecolor=red,urlcolor=blue,bookmarks=false,hypertexnames=true]{hyperref} 
\usepackage{float}
\usepackage{float}
\usepackage{xcolor}
\usepackage{array}
\usepackage{setspace}
\usepackage{lipsum}
\usepackage[
	a4paper,    
	top=1in,    
	bottom=1in, 
	left=1in,   
	right=1in,  
]{geometry}

 \newcommand{\HM}[1]{{\color{black}{#1}}}

\usepackage{hyperref}

\newcolumntype{P}[1]{>{\centering\arraybackslash}p{#1}}
\newcolumntype{M}[1]{>{\centering\arraybackslash}m{#1}}
\usepackage{setspace}
\usepackage{amsmath}
\usepackage{parskip}

\begin{document}
\singlespacing

\title{Dynamics of Transition Metal Ion Transport in High-Gradient Magnetic Fields}

\author{Prateek Benhal}
\affiliation{Department of Chemical and Biomedical Engineering, FAMU-FSU College of Engineering, Tallahassee, FL, 32310, USA}
\affiliation{Center for Rare Earths, Critical Minerals, and Industrial Byproducts, National High Magnetic Field Laboratory, Tallahassee, FL 32310, USA}

\author{Muhammad Garba}
\affiliation{Department of Chemical and Biomedical Engineering, FAMU-FSU College of Engineering, Tallahassee, FL, 32310, USA}
\affiliation{Center for Rare Earths, Critical Minerals, and Industrial Byproducts, National High Magnetic Field Laboratory, Tallahassee, FL 32310, USA}
\author{Jamel Ali}
\affiliation{Department of Chemical and Biomedical Engineering, FAMU-FSU College of Engineering, Tallahassee, FL, 32310, USA}
\affiliation{Center for Rare Earths, Critical Minerals, and Industrial Byproducts, National High Magnetic Field Laboratory, Tallahassee, FL 32310, USA}
\author{Theo Siegrist}
\affiliation{Department of Chemical and Biomedical Engineering, FAMU-FSU College of Engineering, Tallahassee, FL, 32310, USA}
\affiliation{Center for Rare Earths, Critical Minerals, and Industrial Byproducts, National High Magnetic Field Laboratory, Tallahassee, FL 32310, USA}

\author{Munir Humayun}
\affiliation{Center for Rare Earths, Critical Minerals, and Industrial Byproducts, National High Magnetic Field Laboratory, Tallahassee, FL 32310, USA}

\affiliation{Department of Earth, Ocean and Atmospheric Science, Florida State University, Tallahassee, FL 32306, USA.}

\author{Hadi Mohammadigoushki}
\thanks{Corresponding author}\email{hadi.moham@eng.famu.fsu.edu}
\affiliation{Department of Chemical and Biomedical Engineering, FAMU-FSU College of Engineering, Tallahassee, FL, 32310, USA}
\affiliation{Center for Rare Earths, Critical Minerals, and Industrial Byproducts, National High Magnetic Field Laboratory, Tallahassee, FL 32310, USA}
\date{\today}

\begin{abstract}
Magnetic separation has emerged as an eco-friendly and sustainable technique with applications in water purification, chemical separation, biochemical, medical, and mining. In this study, we present, a combined experimental and theoretical investigation of the transport of transition metal ions using high-gradient magnetic fields. Experiments were conducted on aqueous solutions containing either paramagnetic manganese chloride (MnCl$_2$) or diamagnetic zinc chloride (ZnCl$_2$) ions, with concentrations ranging from 1 mM to 100 mM under a non-uniform magnetic field of an electromagnet. Our results demonstrate that while paramagnetic MnCl$_2$ is captured by the mesh wool in the magnetic field, diamagnetic ZnCl$_2$ remains unaffected by the presence of magnetic field. The capture efficiency of paramagnetic MnCl$_2$ increases with both the initial ion concentration and the applied magnetic field strength. Furthermore, in binary mixtures, the capture rate of MnCl$_2$ is reduced compared to single-ion solutions, highlighting the role of ion interactions in magnetic separation. Our theoretical modeling indicates that magnetic capture is governed by a balance between magnetic forces and viscous forces. Additionally, the magnetic separation process is enhanced by the field-induced cluster formation of paramagnetic metal ions, which are predicted to be two orders of magnitude larger than individual hydrated ion units. These findings provide insights into the mechanisms of magnetic transport of metal ions and offer potential pathways for improving separation efficiency in complex ion mixtures that contain critical materials.\par


\end{abstract}
\maketitle
\section{Introduction}


Following the discovery of Faraday on electromagnetism, magnetic separation has long been used as a technique to separate components of a mixture based on their magnetic properties~\cite{gunther1909electro}. This method has found applications in a wide range of processes, from the nuclear industry, drug delivery~\cite{dames2007targeted}, in chemical kinetics~\cite{steiner1989magnetic}, water purification~\cite{yavuz2006low,lim2014challenges} as well as in biochemical and medical fields\cite{kemsheadl1985magnetic,ito2005medical}. As the global economy shifts toward sustainable and renewable energy sources, lithium-ion batteries (LIBs) are increasingly used in nearly every electronic product. With the ever-growing need for LIB production, multimillion tons of LIBs will reach their end of life in the near future and, if not properly recycled, could potentially lead to a detrimental impact on natural resources, supply chain, environment and energy conservation. LIBs, together with a wide range of electronic equipment contain critical metals such as lithium, nickel, cobalt, and manganese, and a key challenge in their recycling is the effective separation of these metal ions from mixed solutions. Current separation methods for these metals are costly, energy intensive, and cause significant environmental pollution\cite{icsildar2019biotechnological,chan2021closed,chan2022separation}. Consequently, there is a pressing need for greener and more environmentally friendly separation techniques. The critical metals found in spent lithium-ion batteries (LIBs) and other electronic equipment display a wide range of magnetic susceptibilities, with some being paramagnetic and others diamagnetic. A paramagnetic material is weakly attracted to a magnetic field, while a diamagnetic material is repelled by a magnetic field. Magnetic separation presents an environmentally sustainable option by reducing waste, improving recycling efficiency, minimizing pollution, and conserving energy~\cite{he2014magnetic,mariani2010high,sierra2013high,qin2021motion}. These advantages make it a promising alternative for recycling critical metals from end-of-life LIBs and other waste streams~\cite{icsildar2019biotechnological}. \par


The conventional magnetic separation devices employ a low gradient magnetic fields (LGMS) and are typically limited to separating strongly magnetic materials, such as iron and magnetite from weakly magnetic materials such as apatite\cite{iranmanesh2017magnetic}. Recognizing that the primary driving force behind magnetic separation is the magnetic field gradient, high-gradient magnetic field separators were invented~\cite{gunther1909electro,sugden1943magnetochemistry,oberteuffer1973high,iranmanesh2017magnetic,kolm1975high}. The main component of high-gradient magnetic field separators involves the presence of a ferromagnetic matrix (in the form of a mesh of wires or grooved plates~\cite{katz1984mat,Mori1992Dev,Gerber1995adv}) in a uniform magnetic field. Over the past fifty years, high-gradient magnetic field separation (HGMS) has attracted considerable interest both experimentally and theoretically. HGMS has typically been successful in efficiently separating the micrometer-scale particles or larger aggregates ~\cite{luborsky1975high,xue2022role,watson1992theoretical,xue2020particle, chen2015high,zheng2017theoretical,zeng2019selective,ye2023separation}. There are a few studies that have shown nano-particles could be captured by the HGMS if they form clusters~\cite{ditsch2005high,moeser2004high} but there is a dearth of HGMS applications to particles from nanometer to atomic scales.\par 

Previous theoretical studies have explored the transport and separation of fine particles in high-gradient magnetic separators (HGMS) through various modeling approaches, including trajectory analysis~\cite{watson1973magnetic,watson1975theory}, stochastic~\cite{svoboda2001realistic} or phenomenological models~\cite{sandulyak1988magnetic,abbasov1999theory,abbasov2002determination}. The first attempt to model the HGMS process was made by Watson in 1973~\cite{watson1973magnetic}. Watson numerically solved the problem of the trajectory of a paramagnetic particle around a ferromagnetic wire by considering inertia and ignoring the effects of viscous and gravitational forces, and developed a relation to assess the change in concentration of fine (typically several micro-meter in size) particles as they exit the HGMS filter~\cite{watson1973magnetic,watson1975theory}. Motivated by some disagreements between the trajectory model and experiments~\cite{arajs1985magnetic}, other researchers developed a filtration
model that takes into account the probability of particle collision and retention on the filter~\cite{svoboda2001realistic}. 
More recently, other researchers have developed phenomenological models~\cite{sandulyak1988magnetic,abbasov1999theory,abbasov2002determination}. However, these models suffer from several simplifying assumptions. For example, these models are developed for static HGMS and do not account for time-dependent capture of particles in the system. In addition, these models do not consider the effects of build-up on the filter. As particles move through the mesh, the ability of the mesh to attract more particles decreases. More recently, Hatton and colleagues developed a model for capturing superparamagnetic nanoparticles, employing a simple force balance that accounts for the effects of viscous forces at vanishingly small Reynolds numbers, particle accumulation under flow and diffusion-limited conditions\cite{ditsch2005high,moeser2004high}. \par 

Despite significant advances in understanding the separation of fine particles (micro- and nano-sized) using high-gradient magnetic fields, a substantial gap remains in determining whether this method can effectively separate metal ions, which are considerably smaller than the finest particles to which HGMS has been successfully applied. At the scale of hydrated metal ions (\HM{in the order of several $\AA$}), in the absence of inertia, diffusive forces become highly significant. Additionally, transition-metal ions, including those found in LIBs, are either paramagnetic or diamagnetic, exhibiting weak magnetizations. Consequently, due to the small size of the ions and their weak magnetic properties, the magnetic force driving their capture by the magnetic mesh is expected to be much weaker than the thermal diffusive forces resisting this capture (see details of analysis in pp. 6-7 of the manuscript). Based on this analysis, at first glance it appears unlikely that transition metal ions with relatively weak magnetic properties could be effectively separated from solution mixtures using high-gradient magnetic separation (HGMS). \HM{Recent experiments showed a detectable enrichment of paramagnetic metal ions in a closed cuvette under LGMS, driven by evaporation~\cite{rodrigues2017magnetomigration,lei2021magnetic,yang2012enrichment}, or in inertial microfluidic flows under  magnetic and/or electric fields~\cite{schroeder2024evaluating}. These findings suggest that additional underlying mechanisms may be at play, warranting further investigation.} \par 
The primary goal of this research paper is to investigate the transport and separation of transition metal ions under high-gradient magnetic fields and to assess the feasibility of separating metal ions from mixtures. To explore this approach, we investigate two transition metal ions: paramagnetic MnCl$_2$ and diamagnetic ZnCl$_2$. We will evaluate the time evolution of the concentrations of these two metal ions in solution with each individual metal ion and solutions that contain a binary mixture of these metal ions. We also present a simple model to predict the evolution of ion concentration within the separation domain, providing insight into the key forces and the underlying physics that govern separation of the metal ions in the high gradient magnetic fields.\par 

\vspace{-0.2cm}
\section{Materials and Methods}
The solutions used in the experiments consist of transition metal ions, specifically manganese (II) chloride (MnCl$_2$) and zinc (II) chloride (ZnCl$_2$) obtained from MilliporeSigma and used as received. The solutions were prepared over a broad range of concentrations from 1mM-100mM in millipore water. Two types of aqueuos solutions are made; solutions containing single metal ions and solutions containing a binary mixture of these metal ions.\par  

\begin{figure}[hbt!]
    \centering
    \includegraphics[width=0.95\linewidth]{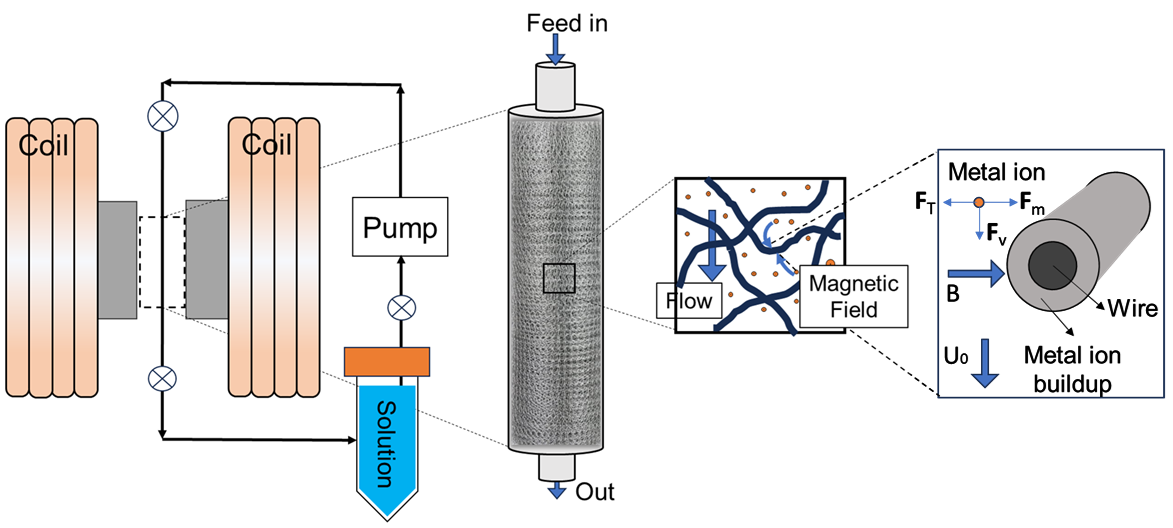}
    \caption{A schematic showing the experimental setup for separation of metal ions. The separation chamber is placed inside a uniform magnetic field between the pole pieces. \HM{Once the flow is initiated, whether in the presence or absence of a magnetic field, we extract 1mL of solution from the solution reservoir (highlighted by blue color) in one hour time interval and measure its concentration.} }
    \label{fig:flowdevice}
\end{figure}

As shown in Fig.~\ref{fig:flowdevice}, the flow chamber consists of several parts. The separation process is conducted under the influence of a 1 Tesla electromagnet, which provided a uniform and controlled
magnetic field during the experiments. The separation chamber is a single-column fabricated from borosilicate glass with a diameter and length of 1cm and 10 cm, respectively. The chamber is filled with a 434-grade steel wool mesh wire with a diameter of 40$\mu$m. \HM{The packing density of the mesh wool is measured to be 0.1 which gives the porosity of 0.9. Using the Carman-Kozeny relation, the permeability of the mesh wool can be estimated as $6.480\times 10^{-10} m^2$}~\cite{schulz2019,rehman2024}. A peristaltic pump is used to generate a continuous flow velocity of approximately 3.3 mm/s, and to circulate the metal-ion solution through a closed-loop system. \HM{A small volume of solution (1 mL) is extracted from the solution reservoir (as shown in Fig.~\ref{fig:flowdevice}) at one-hour intervals for concentration analysis.} To measure the concentration of the metal ions during experiments \HM{(at each one hour interval)}, the light absorbance of the metal ions in the presence of a coloring agent (Xylenone Orange) is measured via a UV-Vis spectrophotometer (3500 Agilent Technologies). Additionally, Inductively Coupled Plasma Optical Emission Spectroscopy (ICP-OES) is used for high-precision ion concentration analysis, particularly in experiments involving binary mixtures of metal ions. The magnetization properties of the materials used in this study were measured by a 16T Quantum Design property measurement system (PPMS) superconducting magnet with a vibrating-sample magnetometer option. \par 


\vspace{-0.4cm}

\section{Modeling Magnetophoretic Separation}

To develop an in-depth understanding of the magnetic separation under high gradients, we first anlyze the key forces at play in this process. In principle, the ions that move through the HGMS may experience a range of forces including magnetic force, gravity, diffusion, viscous force, and inertia. Inertia can be assessed using the Reynolds number, which is defined as: $Re = \frac{\rho |\mathbf{U}_0| d}{\eta}$, where $\rho$, $|\mathbf{U}_0|$, $d$, and $\eta$ are solution density, average fluid velocity imposed by the pump, wire diameter, and the solution viscosity, respectively. The Reynolds number in the experiments reported in this paper is very low $Re<<$1. Therefore, inertia is negligible. \HM{When dissolved in water, metal ions become hydrated by surrounding water molecules, forming a hydration shell. For single hydrated ions, the effective radius of this hypothetical hydration sphere is approximately $\approx$ 3.5 $\r{A}$ \cite{persson2024structure, marcus1988ionic}. However, the hydration sphere may extend to include a partial second hydration shell, potentially incorporating chloride anions. In this case, the effective radius could increase to approximately $R_s \approx 6 $ $\r{A}$. This hydration shell radius is comparable to the hydrodynamic radius of the metal ion. At this scale, the gravitational forces acting on a hypothetical hydration sphere with $ R_s \approx 6 $ $\r{A}$ are negligible.} \HM{On the other hand, under isothermal conditions, the magnetic force that a metal ion with no net charge experiences is related to magnetic energy as\cite{simon2000diamagnetic,parker1990advances,rassolov2024magnetophoresis}:
$\mathbf{F}_m = -\nabla E_{mag}$. Magnetic energy is expressed as: $E_{mag} = -\mu_0 V_s \mathbf{M_s}\mathbf{H}/2$, where $\mu_0$ is the permeability of free space $(4\pi \times 10^{-7}  \frac{N}{A^{2}})$, $V_s$ is the volume of the metal ion, $\mathbf{M_s}$ is the magnetization of the metal ion, and $\mathbf{H}$ is the magnetic field strength created around the ion by the presence of the wire. Therefore, the magnetic force can be expressed as:
\begin{equation}
\mathbf{F}_m = \mu_0 V_s (\mathbf{M_s} \cdot \nabla) \mathbf{H} + \frac{\mu_0 V_s \mathbf{M_s} \mathbf{H} \cdot \nabla c}{2c}.
\label{magforce}
\end{equation}
The first term on the right-hand side of Eq.~(\ref{magforce}) represents the magnetic force due to field gradients ($ F_{\nabla \mathbf{B}}$), while the second term accounts for the contribution from concentration gradients ($ F_{\nabla \mathbf{c}}$). Our calculations show that $F_{\nabla \mathbf{c}}$ is significantly smaller than $ F_{\nabla \mathbf{B}}$ in the experiments of this study (see the right panel of Fig.~A(1) in the supplementary materials and the related discussion). Therefore, throughout the rest of this manuscript, we focus exclusively on the magnetic force arising from field gradients.} The magnetic field around the wire depends on the magnetization of the wire $\mathbf{M_w}$ and the relative distance from the surface of the wire. In addition to the magnetic force, the metal ion may experience thermal diffusion as a result of Brownian motion. This force could be approximated as: $\mathbf{F}_{T} \approx {k_B T}/{R_s}$, where $k_B$, and $T$ are the Boltzmann constant, and the absolute temperature, respectively. Finally, in a flowing system, ions will experience a hydrodynamic force that in the absence of inertia could be estimated as: $\mathbf{F}_{v} \approx 6\pi \eta \mathbf{U}_{0} R$. \par 

In a multi-collector magnetic filter such as a steel wool separation matrix used in this study, and for a paramagnetic ion, the magnetic force is attractive towards the wire, whereas, the diffusive and viscous forces are competing against this attractive force. To assess the relative importance of these forces, we can construct two dimensionless numbers. First, the ratio of the magnetic force to thermal diffusion is expressed as a magnetic Peclet number:
\begin{equation}
    \mathrm{Pe_m} = \frac{\mathbf{F_m}}{\mathbf{F_T}}= \frac{\mu_0 V_s (\mathbf{M_s} \cdot \nabla) \mathbf{H}}{k_B T/R_s}.
    \label{eqn:Peclet-definition}
\end{equation}
Note that the magnetic Peclet number varies as a function of the relative distance from the wire surface. In the vicinity of the wire, where the magnetic field gradient is at its peak, the magnetic Peclet number also reaches its maximum. Previous studies have calcutated the maximum magnetic field around a magnetic wire~\cite{moeser2004high,ditsch2005high,fletcher1991fine}, which can be used to construct the maximum magnetic Peclet number as: 
\begin{equation}
\mathrm{Pe_m}|_{max} = \frac{4\pi\mu_o M_{wire} M_{s}R_{s}^3}{3k_{B}T}.
\label{eqn:Peclet-max}
\end{equation}
Here $M_{wire}$ is the magnetization of the wire. Secondly, the relative importance of the magnetic force and the viscous forces in the vicinity of the wire, can be estimated as a Mason number defined below: 
\begin{equation}
\mathrm{Ma_m}|_{max} = \frac{\mathbf{F_m}}{\mathbf{F_v}}|_{max} = \frac{2\mu_o M_{wire} M_{s}R_{s}}{9\eta |\mathbf{U}_0|}.
\label{eqn:Mason}
\end{equation}
Depending on the relative importance of these forces, the diffusion or flow could be the rate limiting factor in magnetically separating and capturing the metal ions. \par

While the above analysis is crucial, it does not enable us to evaluate the temporal variation of ion concentration within the column. To address this, we will need to apply a mass conservation approach, as outlined below. Because inertia is not important in this work, the HGMS model developed by Watson\cite{watson1973magnetic} can not be applied to the experiments. As paramagnetic metal ions move from the top of the column to the bottom and exit the column, a portion of metal ions is captured by the wires. To evaluate the capture rate of these species in this system, one could write the mass conservation on the species as follows: 
\begin{equation}\label{eqn_mass}
    \frac{\partial n}{\partial t} = - \nabla \cdot (D\nabla n) + \nabla \cdot (n\mathbf{v}).
\end{equation}
Here $n$, $D$ and $\mathbf{v}$ are the number density of species, diffusion coefficient and the fluid velocity. Under steady state conditions, and constant imposed fluid velocity, the mass balance along the separation column height can be written as: 
\begin{equation}\label{eqnn}
    \frac{\partial n}{\partial x}=\alpha{n}.
\end{equation}
Previous studies showed that $\alpha = -f/L$\cite{moeser2004high,fletcher1991fine, ditsch2005high}, where $f$ and $L$ are the fractional capture of species in the column and the length of the column, respectively. The fractional capture of species $f$ is given  by the expression below\cite{moeser2004high,ditsch2005high}:
 \begin{equation}\label{eqnC}
     f=\frac{A_c-b}{\phi_i-b}
 \end{equation}
Here $A_c = {A(1-\phi_i)}/{2\pi}$ and $\phi_i$, $b$ are the capture area of species around the wire, the initial void fraction of the column, and the buildup of species in the column, respectively. As species move through the column, they accumulate, and the rate of this buildup—impacting the capture efficiency of species by the wire—has been calculated to be\cite{moeser2004high,fletcher1991fine, ditsch2005high}:
\begin{equation}\label{eqnB}
    \frac{\partial b}{\partial t}=-\frac{n f}{n_s L}|\mathbf{U_o}|
\end{equation}
Here $n_s$ is the number density of species that are captured on the wire. In what follows, we adopt the above model to predict the variation of ion concentration in our experiments. Note that we calculate the concentration of effluent after each filtration pass by simultaneously solving Eqs.~(\ref{eqn_mass}-\ref{eqnB}). We will then use the existing buildup in the column to assess the variation of ion concentration for subsequent filteration passes. The details of the equations and computational results for capture rate under diffusion and flow conditions are discussed in detail in the supplementary materials.\par

\section{Results and Discussion}

\subsection{Force Balance Analysis}

\begin{figure}[hthp]
    \centering
    \includegraphics[width=0.292\linewidth]{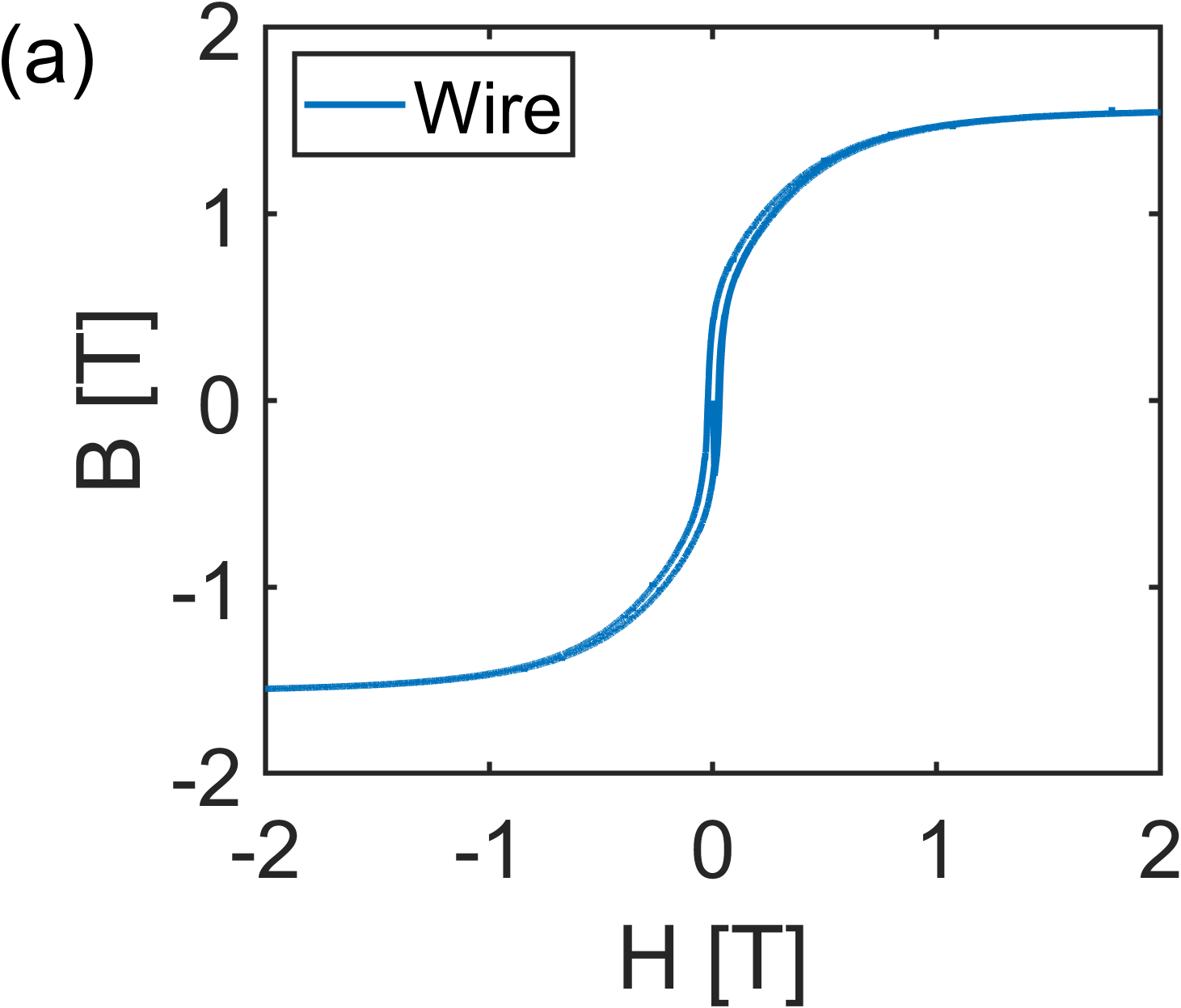}
    \includegraphics[width=0.342\linewidth]{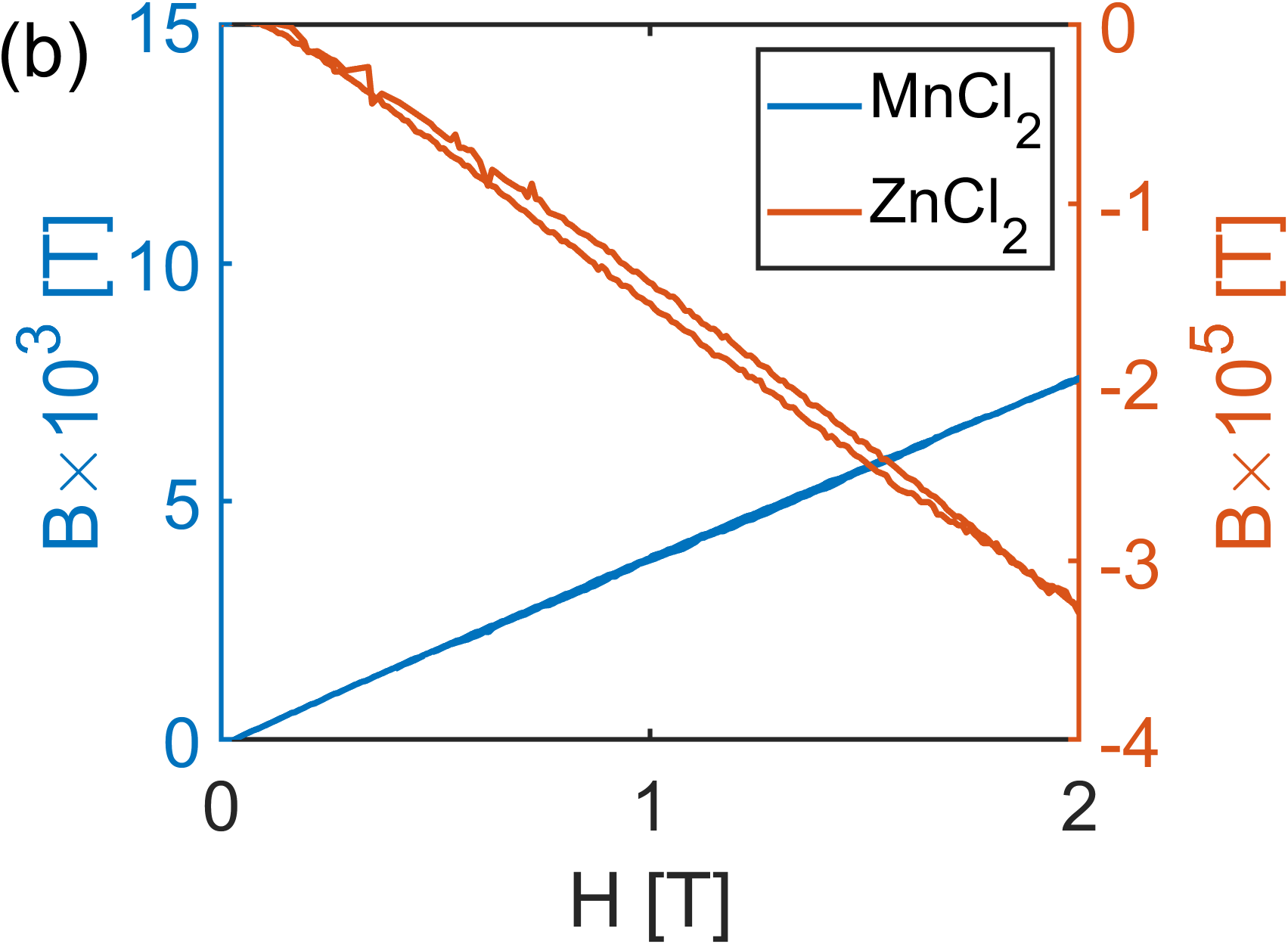}
    \includegraphics[width=0.33\linewidth]{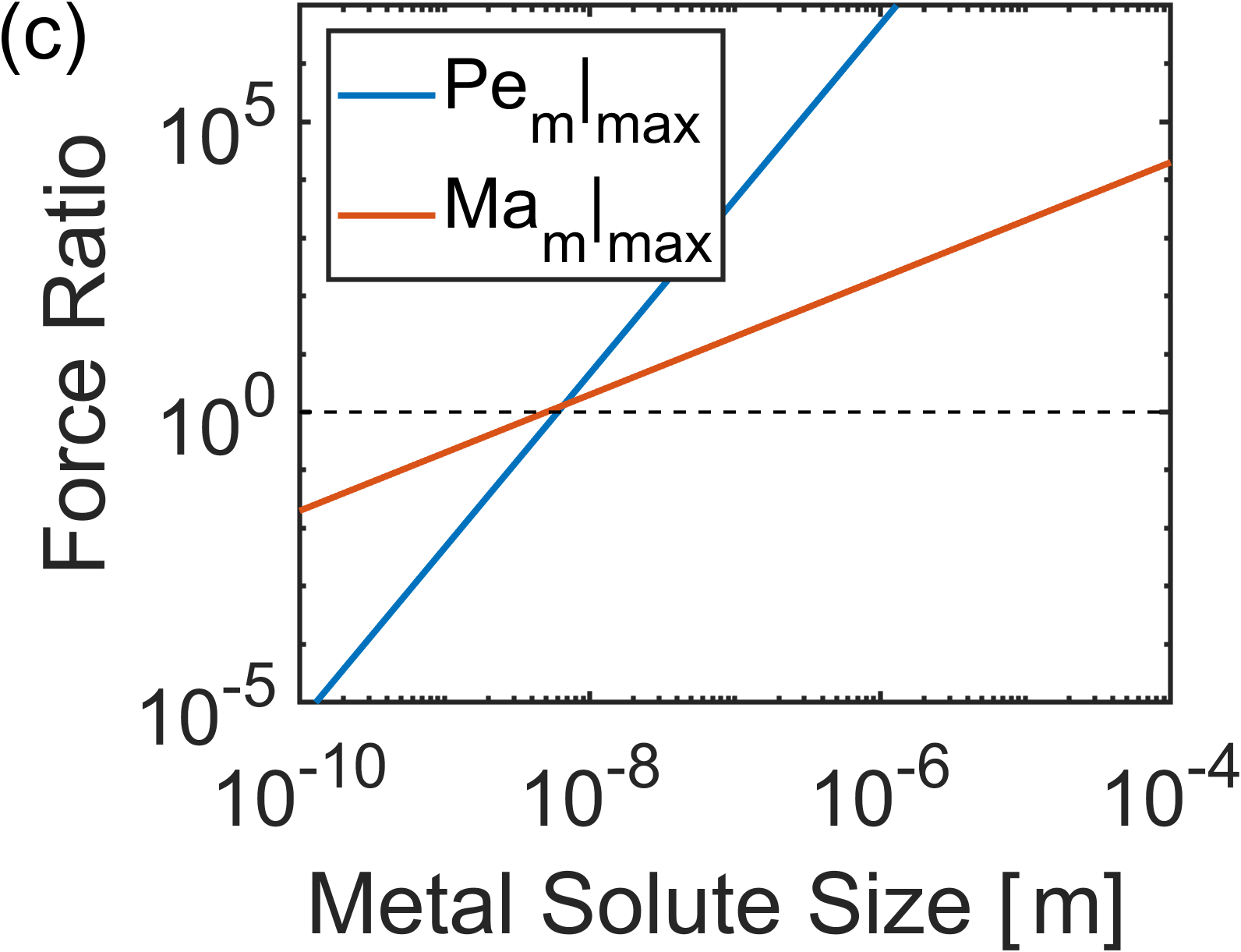}
    \caption{Magnetization curves measured for (a) wire mesh, (b) MnCl$_2$ and ZnCl$_2$ ions. (c) The calculate ratio of forces noted in Eq.~(\ref{eqn:Peclet-max},\ref{eqn:Mason}) as a function of paramagnetic MnCl$_2$ ion size for the experimental conditions used in this study and a magnetic field of 1 T. Below a critical size of $R_c\approx$ 8nm, the separation process of the metal ions is diffusion limited (diffusion is the rate limiting factor). Beyond this critical size, the separation is limited by the viscous forces.}
    \label{fig:Forceratio}
\end{figure}

Prior to discussing the experimental results, we begin our analysis by evaluating the dimensionless numbers. To evaluate the dimensionless paramaters, we will need magnetization of the wire and the metal ions. Fig.~\ref{fig:Forceratio}(a,b) presents the magnetization curves of stainless steel mesh wool, as well as those of paramagnetic and diamagnetic samples. While the magnetization of the wire saturates for applied magnetic fields above 1T, the metal ions do not reach saturation within this range of magnetic fields. Using the experimentally determined magnetization of the wire, and those of MnCl$_2$, we have calculated the ratio of these two forces as a function of ion size for an applied magnetic field of 1T, as shown in Fig.~\ref{fig:Forceratio}(c). Our calculations indicate that below a critical ion size of approximately 8 nm, diffusive forces dominate, while beyond this threshold, magnetic separation is limited by viscous forces. This simple dimensionless analysis suggests that the ion size must exceed this critical threshold of 8 nm for any meaningful separation of the paramagnetic metal salt from their solution. Notably, this size is significantly larger than the hydration radius of individual metal ions. \HM{As noted before, the typical hydration radius for MnCl$_2$ complexes in water is approximately 6 $\r{A}$ \cite{persson2024structure,marcus1988ionic}}. Under this condition, both $\mathrm{Pe_m}|_{max}$ and $\mathrm{Ma_m}|_{max}$ are much smaller than unity suggesting that no magnetically assisted capture should occur for paramagnetic metal ions. In addition, the magnetic force for the diamagnetic ZnCl$_2$ is always negative, indicating that these ions are repelled by the mesh wire. As a result of this force balance, it would be expected that neither paramagnetic nor diamagnetic ions would be captured by the mesh wool. Surprisingly, our experiments demonstrate that magnetic capture can occur for paramagnetic metal ions under certain conditions. In the following sections, we will delve into the experimental results and present a comprehensive analysis of our findings, highlighting the mechanisms behind this unexpected phenomenon.\par

\subsection{Magnetophoretic Separation of Individual Metal ions from Solution}

\begin{figure}[hthp]
    \centering
    \includegraphics[width=0.4\linewidth]{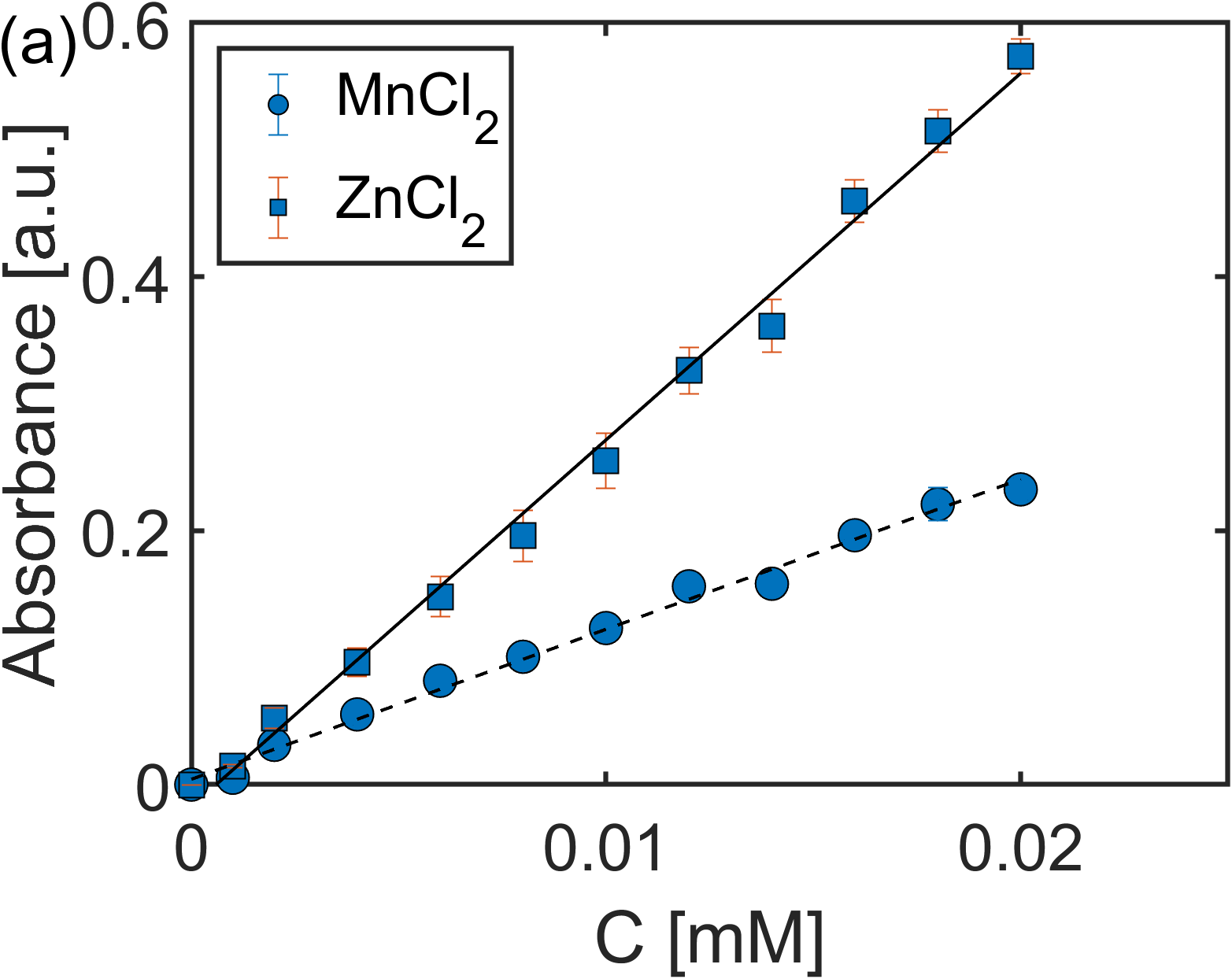}
    \includegraphics[width=0.4\linewidth]{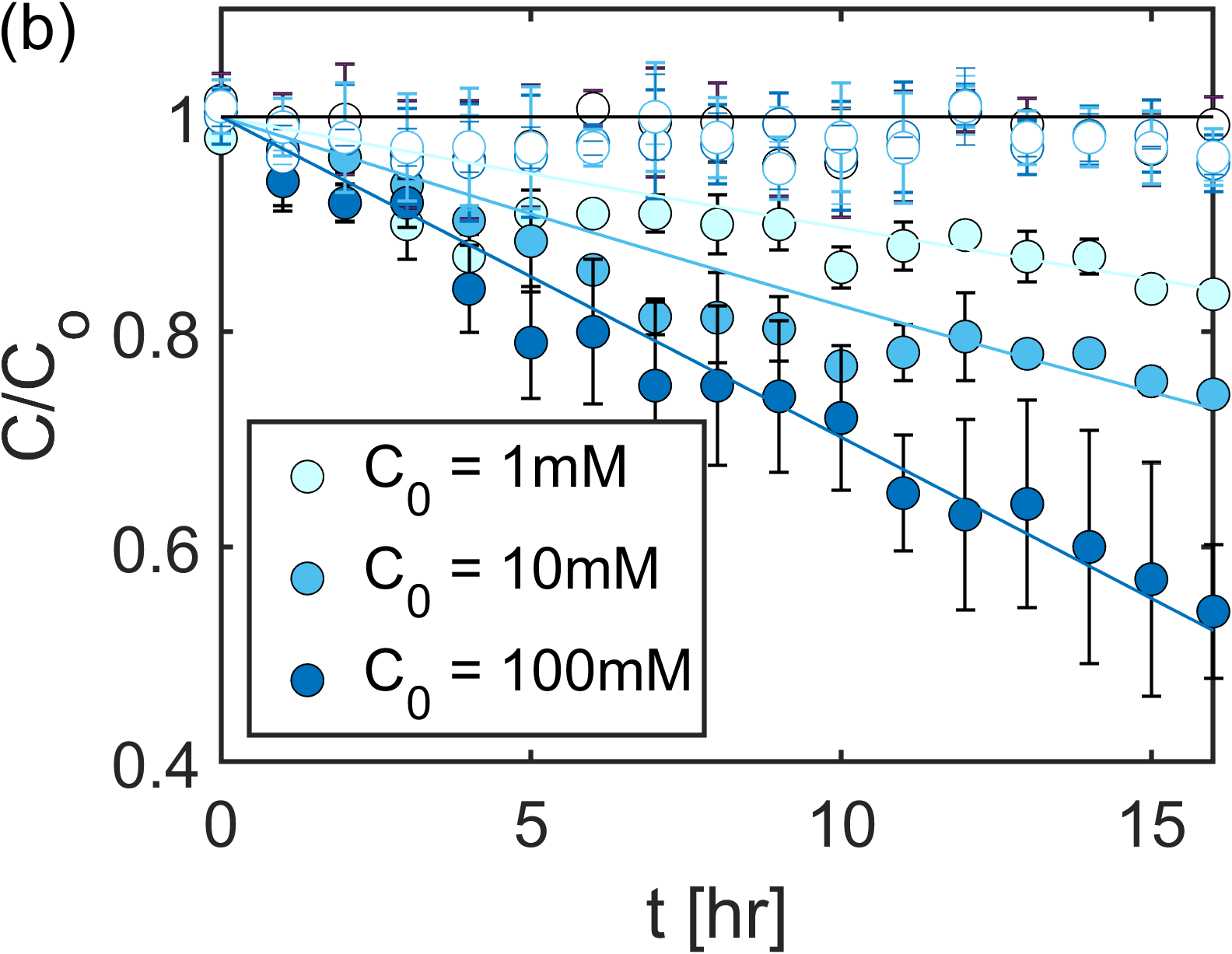}
    \includegraphics[width=0.4\linewidth]{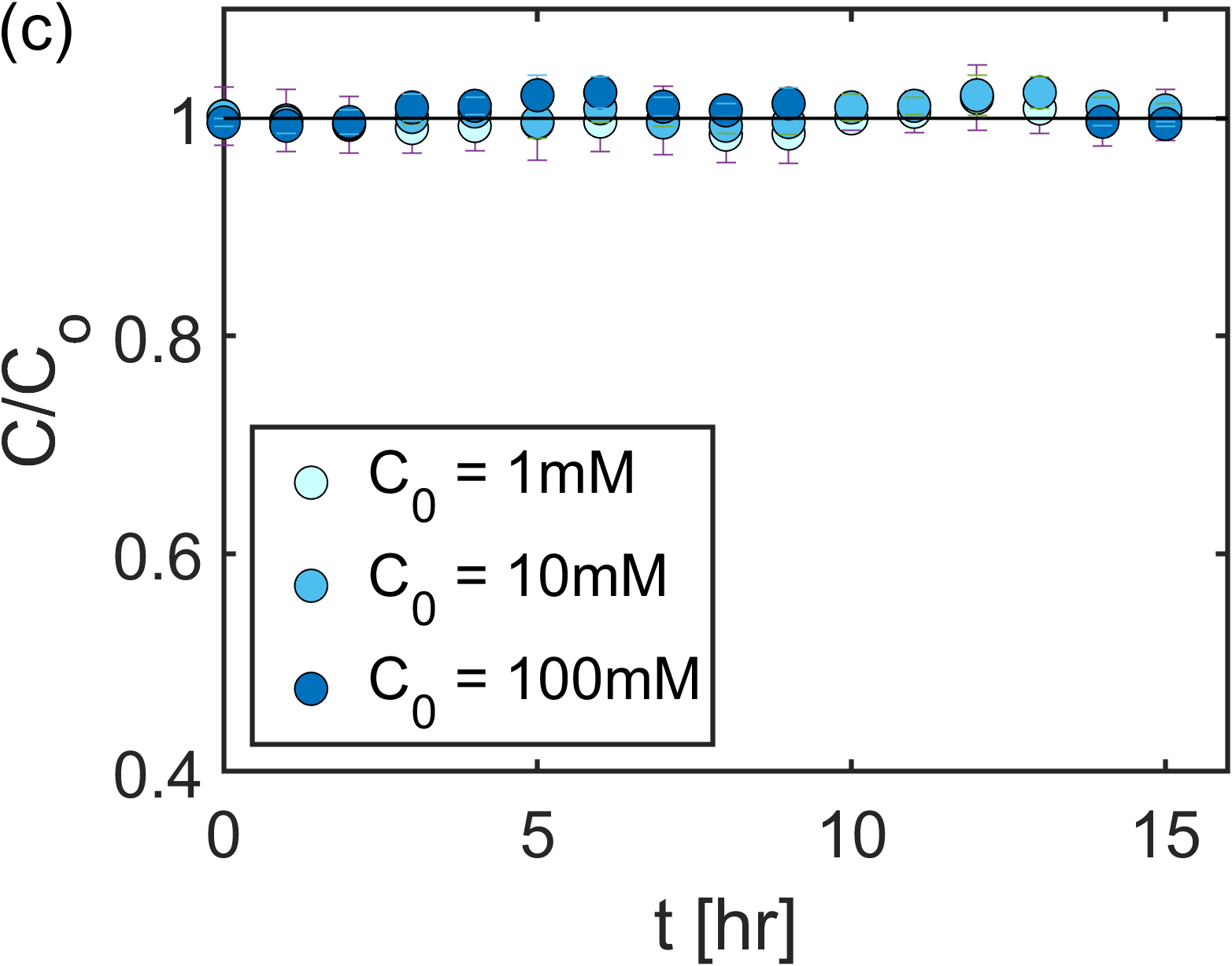}    \includegraphics[width=0.4\linewidth]{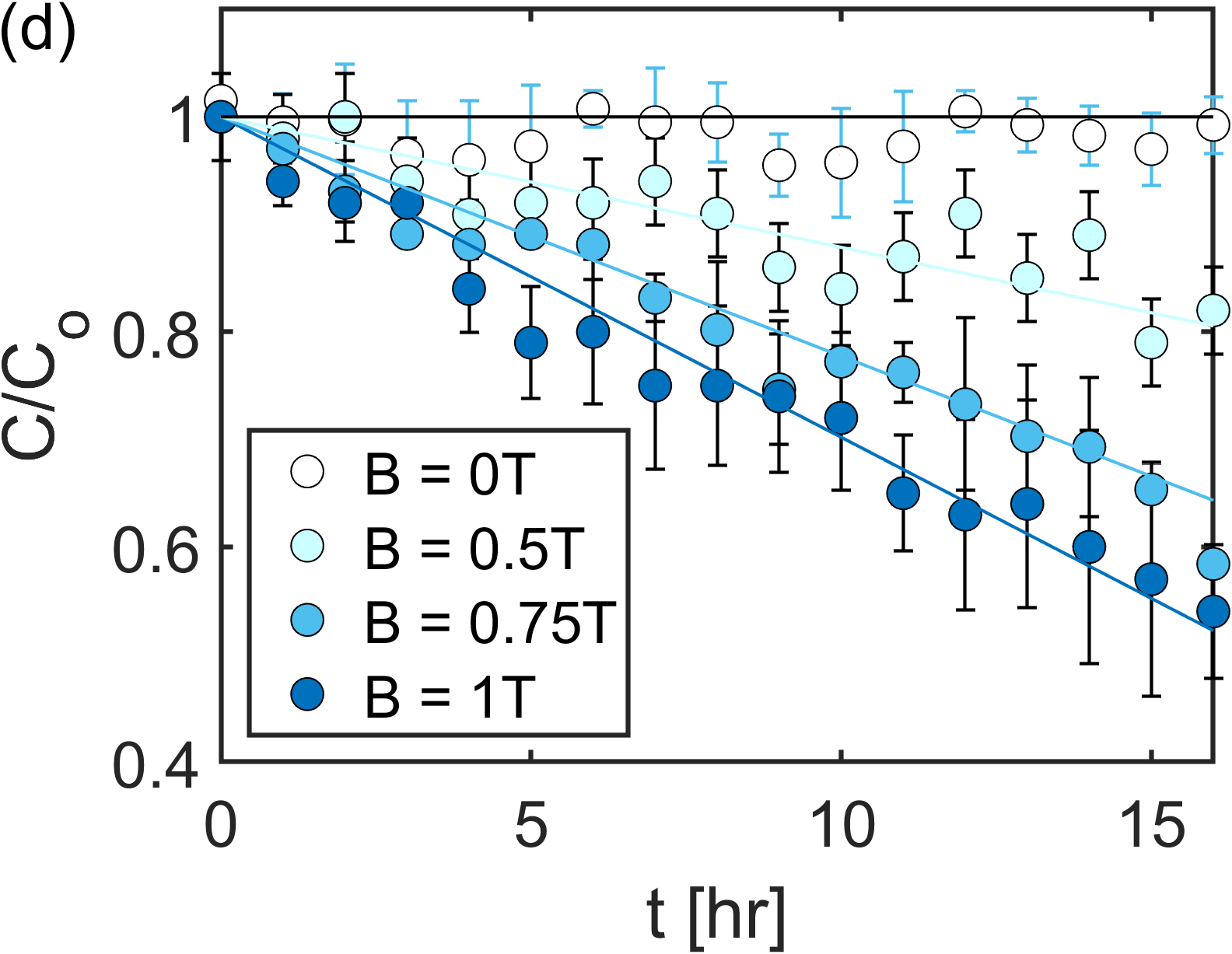}
    \includegraphics[width=0.4\linewidth]{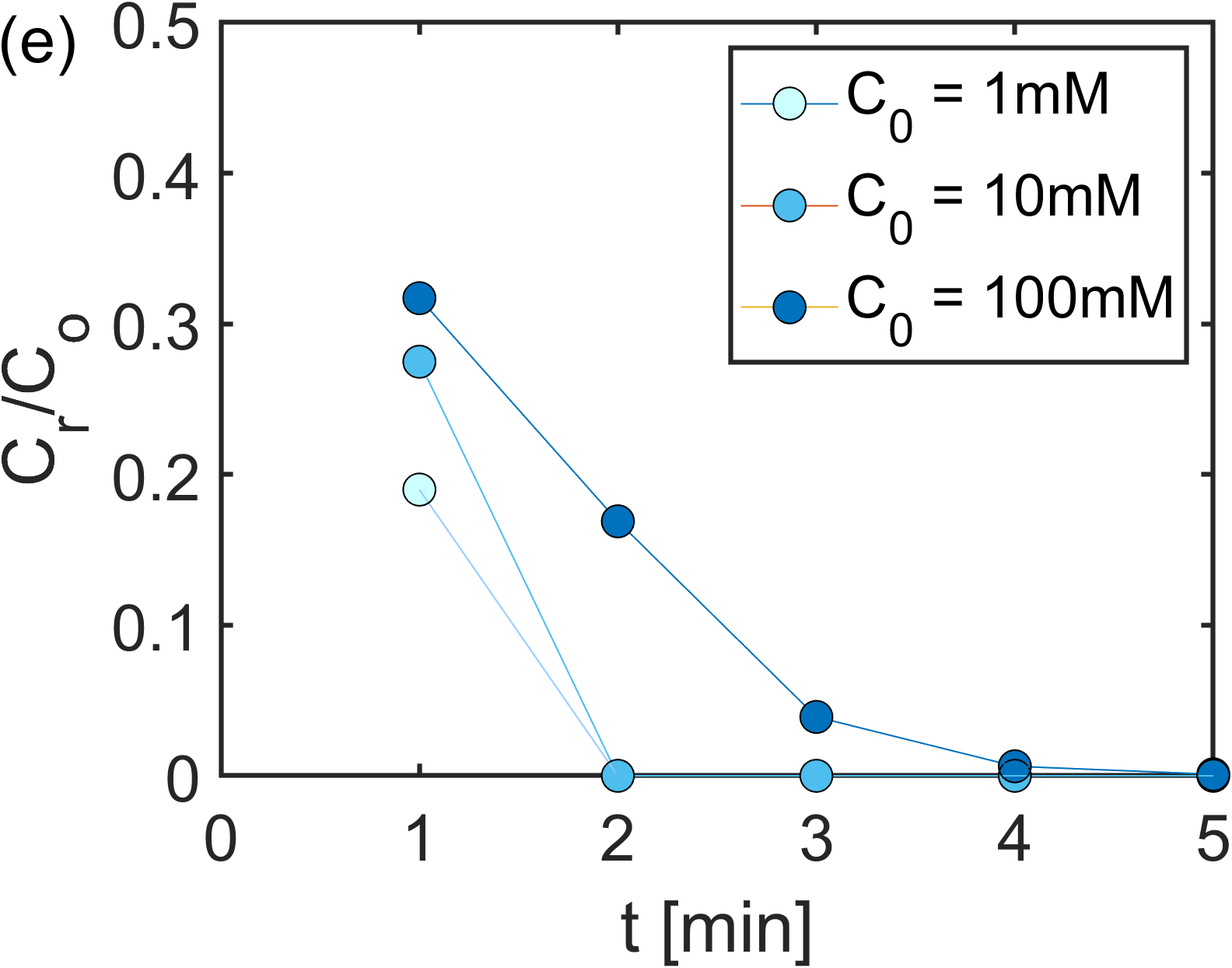}
    \includegraphics[width=0.4\linewidth]{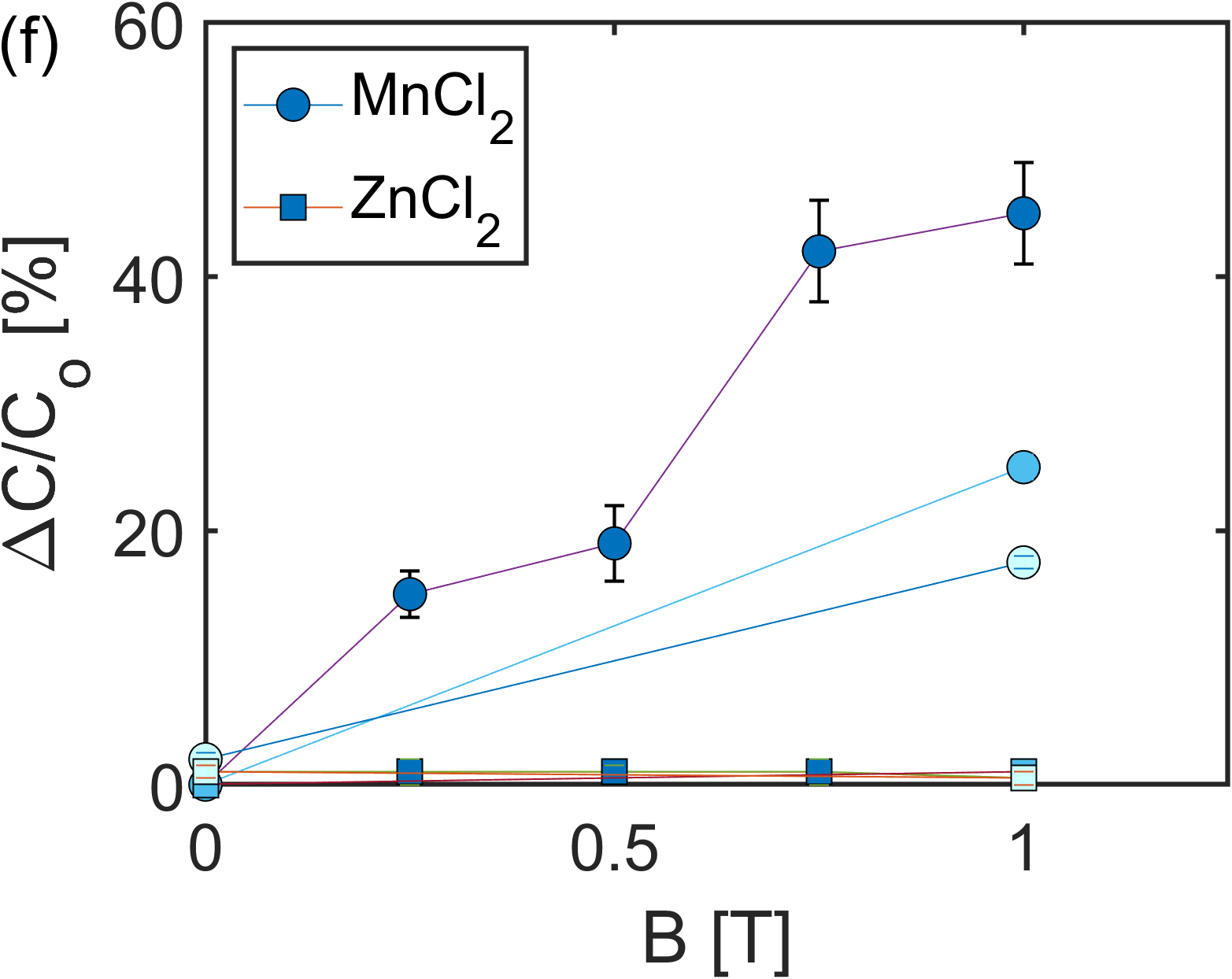}
    \caption{(a) The calibration curves measured for both metal ions. The slope of these linear lines that are best fitted to the experimental data are: 11.82 and 28.88 for MnCl$_2$ and ZnCl$_2$ respectively. (b) Normalized concentration of captured MnCl$_2$ metal ions as a function of time for a broad range of initial concentrations. The filled and open symbols correspond to magnetic field of 1 T and 0 T, respectively. (c) Normalized concentration of ZnCl$_2$ metal ions as a function of time for a broad range of initial concentrations at at magnetic field of 1 T. (d) The normalized captured concentration of MnCl$_2$ as a function of time for various imposed magnetic fields and $C_0 =$ 100mM. (e) Normalized recovered MnCl$_2$ concentration as a function of flushed time for various initial concentrations after field was turned off from 1T. (f) Overall change in concentration of the metal ions as a function of applied magnetic field after 16hrs of experiments. The light to dark colors indicate $C_0 =$1\HM{mM}, 10\HM{mM} and 100mM, respectively.}
    \label{fig:singleelements}
\end{figure}

We began experiments by flowing solutions containing single metal ions in deionized water through the column. To ensure there were no complications arising from interactions between the stainless-steel mesh wool and the aqueous solutions, we conducted control experiments in parallel, where the magnetic field was kept off.  Fig.~\ref{fig:singleelements}(a) shows the absorbance intensity for these metal ion solutions as a function of concentration. Within this concentration range, the absorbance intensity increases linearly with the concentration of the ion, indicating that the Beer-Lambert law is applicable\cite{swinehart1962beer}. This plot serves as a calibration curve to determine the concentration of metal ions extracted from the column. \par 

Fig.~\ref{fig:singleelements}(b) illustrates the temporal evolution of the normalized concentration of paramagnetic MnCl$_2$ for different initial concentrations. The paramagnetic metal ions are captured by the mesh wool, with the capture rate increasing over time. Furthermore, as the initial concentration of MnCl$_2$ increases, \HM{the amount of captured metal ion increases (or normalized concentration of metal ions in the running solution decreases)}. It is important to note that control experiments, where the magnetic field was turned off, showed no significant change in the concentration of paramagnetic MnCl$_2$ ions over time. This indicates that the observed capture is directly related to the presence of the magnetic field. Fig.~\ref{fig:singleelements}(c) shows the temporal variation of the normalized concentration for diamagnetic ZnCl$_2$ ions across a wide range of initial concentrations. The concentration remains unchanged, even after 16 hours. Unlike paramagnetic ions, the diamagnetic ZnCl$_2$ ions do not interact with the magnetic mesh wool and, therefore, are not captured by the stainless-steel mesh. Diamagnetic ZnCl$_2$ ions are expected to experience a repulsive force from the magnetic mesh, since their magnetic susceptibility is less than that of deionized water. Such an effect is not expected to impact the output of ZnCl$_2$ ions as evidenced by Fig.~\ref{fig:singleelements}(c). In addition, the effect of varying external magnetic field strengths on the capture rate of paramagnetic MnCl$_2$ was studied. Fig.~\ref{fig:singleelements}(d) illustrates the normalized concentration of MnCl$_2$ as a function of time for different magnetic field intensities. As the imposed magnetic field strength increases, the overall rate of ion capture rises correspondingly. These results support our hypothesis that the separation process is induced by the influence of the external magnetic field.  \par  

To further confirm this hypothesis, we turned off the magnetic field following the experiments and subsequently flushed the chamber with purified water. The concentration of the metal ion in the flushed fluid was measured at 1-minute time intervals. Fig.~\ref{fig:singleelements}(e) presents the normalized concentration of recovered metal ions as a function of time for three distinct initial concentrations, following experiments conducted at a magnetic field strength of $B=$1T. The data demonstrate that paramagnetic ions can be recovered from the chamber upon deactivation of the magnetic field, thereby reinforcing our hypothesis that the separation process is driven by the external magnetic field. Furthermore, the results underscore the reversible nature of magnetic capture in these systems. Finally, Fig.~\ref{fig:singleelements}(f) summarizes the capture rate of the two metal ions after 16 hours of magnetic field exposure, plotted against the applied magnetic field across the range of initial concentrations tested in this study. The results clearly indicate that magnetic separation is enhanced with increasing magnetic field strength, with higher initial concentrations yielding greater separation efficiency for the paramagnetic MnCl$_2$ ions. In contrast, the diamagnetic ZnCl$_2$ ions show no interaction with the magnetic field, regardless of its strength or the initial ion concentration, and thus, their concentration remains unchanged throughout the experiments.\par 

As indicated by the force balance analysis, magnetically assisted capture was not anticipated for paramagnetic metal ions. This raises the question of what might have contributed to this discrepancy with our experimental results. The direct measurement of the forces acting on metal ions presents significant challenges. However, we can gain further insight into this issue by examining the theoretically predicted temporal evolution of metal ion concentration and comparing it with our experimentally measured values. This comparison could help elucidate the underlying factors influencing the observed magnetic capture. In our first attempt to calculate the concentration variation in the domain, we solved Eq.~(\ref{eqn_mass}-\ref{eqnB}) by using the typical hydration radii of the metal ions in solution as a reference (e.g., approximately 6 $\r{A}$ for MnCl$_2$ and ZnCl$_2$\cite{persson2024structure,marcus1988ionic}). Under this condition, the paramagnetic MnCl$_2$ shows no interaction with the magnetic field, as indicated by the normalized concentration remaining around unity, suggesting no magnetic capture (see the horizontal line in Fig.~\ref{fig:singleelements}(b)). This result aligns with the previously presented force balance analysis.\par 

To find the best match between experiments and the model, \HM{we increased $R_s$ values beyond the hydration radius of the metal ion} to obtain the best fit with the experimental data. The continuous lines in Fig.~\ref{fig:singleelements}(b) represent the best fit of the model to the experimental data points. Furthermore, the model was fitted to the experimental results obtained at different magnetic field strengths, as shown in Fig.~\ref{fig:singleelements}(d). Table~(\ref{Table:ModelingParameters}) provides a summary of the ion radii that yielded the best fit with the experimental data. \HM{Several observations can be made here. First, it is noteworthy that the effective ion size that correlates best with the experimental results is substantially larger, by nearly two orders of magnitude, than the hydration radius of the individual metal ions. This observation suggests that the paramagnetic metal ions may have aggregated into clusters under the influence of the magnetic field, leading to their enhanced capture rate in the experiments. }
\begin{table}[hthp]
 \centering
    \begin{tabular}{c@{\hskip 0.5cm}c@{\hskip 0.5cm}c@{\hspace{0.5cm}}c@{\hspace{0.5cm}}c@{\hspace{0.2cm}}}
        \hline\hline
     Solution & ion & $c_0$ [mM] & $\mathbf{B}$ [T] & $R_s$ [nm]  \\ \hline\hline
       &      & 1 & 1 &  
              15.4      \\
       &      & 10 & 1 & 13.8     \\
   Single & MnCl$_2$ & 100 & 1 & 11.8    \\
      &       & 100 & 0.75 & 13.2 \\
      &       & 100 & 0.5 & 15.8  \\
     &        & 100 & 0.25 & 22.1 \\
             
      \hline

    \end{tabular}
    \caption{The model parameters used to find the best match with experimental data measured in the separation chamber.}
    \label{Table:ModelingParameters}
\end{table}
Previous studies have demonstrated that transition metal ions in porous media experience a pronounced magnetophoresis effect, much stronger than predicted by a force balance in individual metal ions\cite{Fujiwara2004JPhysChemB,Fujiwara2006JPhysChemB,Franczak2016PhysChemChemPhys}. Researchers have hypothesized that this discrepancy may be due to the formation of ion clusters under the influence of a non-uniform magnetic field. This hypothesis aligns with the findings from our experimental and modeling efforts discussed above. \HM{The concept of magnetic field-induced cluster formation is well-established and has been extensively studied in colloidal systems and suspensions of superparamagnetic nanoparticles through both experimental investigations and theoretical modeling~\cite{tsouris1995flocculation,faraudo2016predicting,faraudo2013understanding,andreu2011aggregation,liu1995field,de2008low}. Prior research has explored the fundamental mechanisms governing cluster formation, including the interplay between particles dipole-dipole interactions, electrostatic repulsion, and van der Waals forces under uniform applied magnetic fields~\cite{tsouris1995flocculation,faraudo2016predicting,faraudo2013understanding,andreu2011aggregation,liu1995field,de2008low}. Under a uniform magnetic field, two fundamental conditions must be satisfied for field-induced particle clustering to occur. First, magnetic forces, primarily governed by dipole-dipole interactions, must be strong enough to overcome random thermal motion. Second, the particle concentration must be sufficiently high to ensure that the rate of aggregation surpasses the rate of dissociation. The strength of these magnetic interactions can be effectively characterized by the magnetic coupling parameter, $\Gamma$, which serves as a key determinant of the system's clustering behavior. For paramagnetic particles, $\Gamma$ is defined as~\cite{liu1995field}:\begin{equation}
\Gamma = \frac{\pi R_p^3 \Delta \chi^2 \mathbf{B}^2}{9 \mu_0 k_B T},
\label{eq:mag-coupling-param}
\end{equation}
Here $\Delta \chi$ is the difference between magnetic susceptibility of the metal ion and the solvent, $R_p$ is the particle radius, and $\mathbf{B}$ is the magnetic flux density. The latter condition for field-induced particle cluster formation is quantified through the parameter $N^*$, which is given by:
\begin{equation}
N^* = \sqrt{\frac{c_0}{\rho_p} \exp{(\Gamma - 1)}}.
\end{equation}
Here $\rho_p$ is the density of the metal ion in the solution. Field-induced aggregation is expected only when \(\Gamma > 1\) and \(N^* > 1\). Fig.~(\ref{fig:cluster}) below illustrates the variation of \(N^*\) and \(\Gamma\) as a function of magnetic field strength and $R_p$ for paramagnetic metal ions of MnCl$_2$. For an initial concentration of \(c_0 = 100\) mM, the analysis predicts field-induced aggregation only for \(R_p > 220\) nm and \(\mathbf{B} \geq 1\) T. According to this analysis, the effective particle (or cluster) size that is required for formation of field-induced cluster formation is much larger than the values predicted by our 1D numerical simulations of the HGMS (see data in Table~(\ref{Table:ModelingParameters})). }\par 

\begin{figure}[hbt!]
    \centering
    \includegraphics[width=0.9\linewidth]{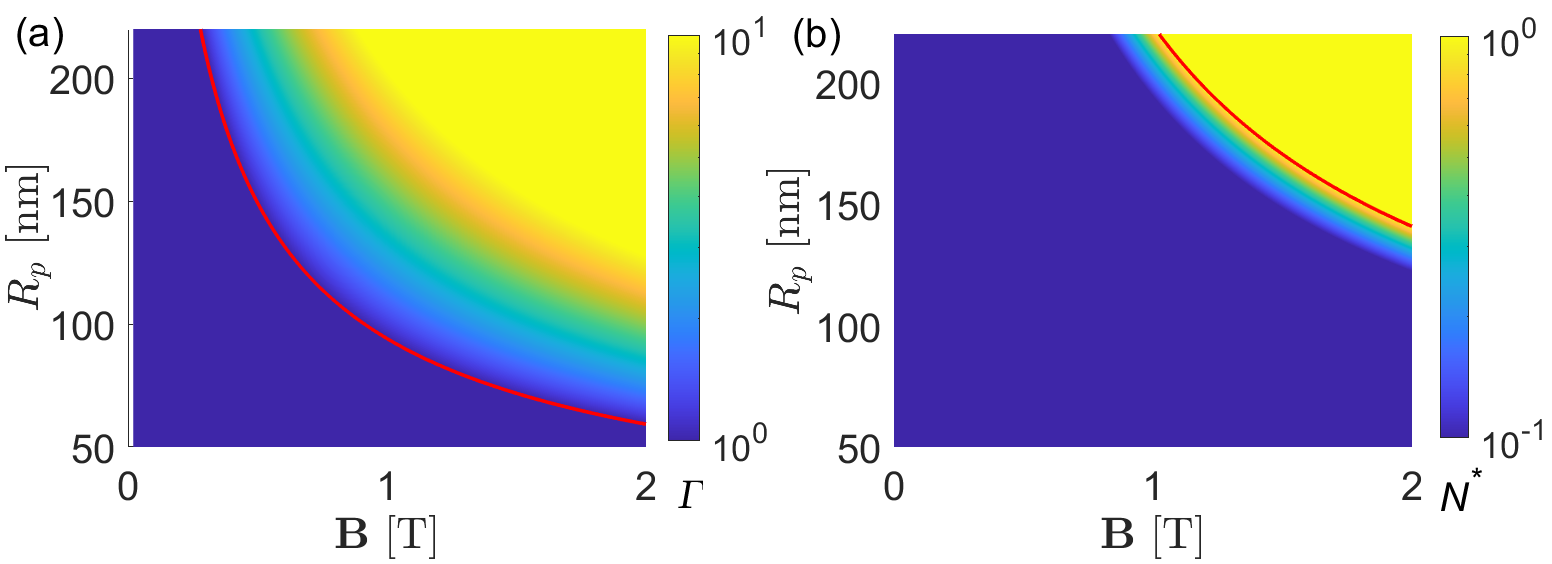}
    \caption{2D heat map of dimensionless parameters $\Gamma$ (a), and $N^*$ (b). The red curve shows $\Gamma =1$ and $N^* = 1$ for each subplot. }
    \label{fig:cluster}
\end{figure}

\HM{This discrepancy is not unexpected. It is crucial to recognize that the above analysis, based on $\Gamma$ and $N^*$, applies strictly to uniform magnetic fields and does not account for the influence of strong magnetic field gradients. In the presence of field gradients, a significant magnetophoretic force may arise, effectively driving metal ions toward the wire and facilitating their mutual attraction. In our study, the magnetic field gradients around individual wires are exceptionally high. Analytical solution to the magnetic field around a ferromagnetic wire~\cite{svoboda2004magnetic} indicates that the magnetic field gradients near the wire surface can reach $(\mathbf{B} \cdot \nabla) \mathbf{B}|_\mathrm{max} \approx $\(10^5\) T\(^2\)/m and above (see left panel of Fig.~(A1) in the supplementary materials.), an exceptionally strong gradient capable of inducing metal ion migration toward the wire and potentially facilitating cluster formation. As demonstrated in our analysis of Fig.~(\ref{fig:Forceratio}), the magnetophoretic force can surpass thermal diffusion (i.e., Pe$_m > 1$) for  $R_s \approx 6$ nm, and the hydrodynamic force (i.e., Ma$_m > 1$) for $R_s \approx 5$ nm, highlighting the significant role of magnetic field gradients in governing metal ion dynamics. Hence, future investigations into field-induced cluster formation should refine the interaction potential between particles to incorporate the effects of magnetophoretic forces under non-uniform magnetic fields. This refinement will enable a more comprehensive understanding of cluster formation mechanisms, particularly in the presence of strong magnetic field gradients~\cite{tsouris1995flocculation}}.\par

\begin{figure}[hbt!]
    \centering
\includegraphics[width=0.5\linewidth]{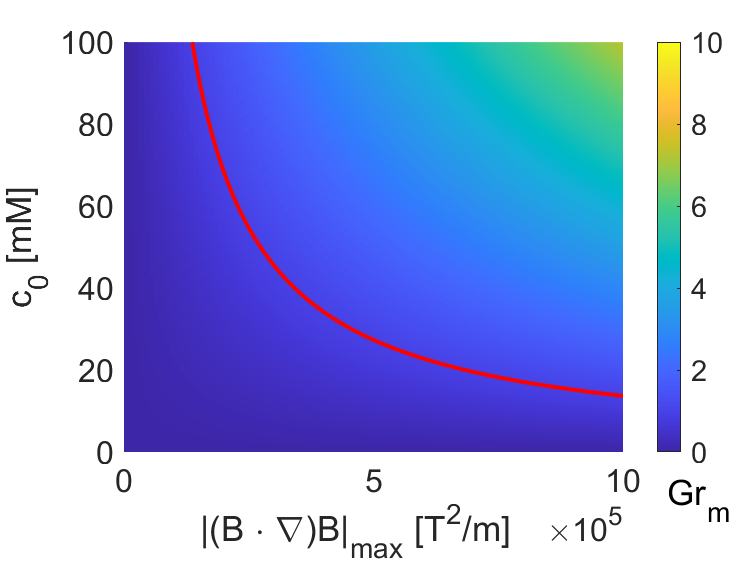}
    \caption{Variation of the magnetic Grashof number with $(\mathbf{B} \cdot \nabla) \mathbf{B}|_\mathrm{max}$ and initial concentration of paramagnetic metal ions of MnCl$_2$ in the solution. The red line marks $\mathrm{Gr}_m = 1$. }
    \label{fig:grashof}
\end{figure}

\HM{Additionally, the data in Table~(\ref{Table:ModelingParameters}) indicate that at a fixed external magnetic field strength ($\mathbf{B} = 1$ T), an increase in the initial metal ion concentration results in a slight decrease in the minimum cluster size ($R_s$) required to align simulations with experiments. Furthermore, as the magnetic field strength increases, $R_s$ decreases even further. These findings are somewhat unexpected, as one would anticipate, based on the analysis presented in Fig.~(\ref{fig:cluster}), that higher magnetic field strengths and initial concentrations would promote the formation of larger clusters (or lead to larger effective cluster sizes). It is important to highlight that, in the discussion above, we have not yet considered the potential formation of secondary flows in the HGMS near wire surfaces. In regions where magnetic forces are strong, such as near the wire surface, the transport of metal ions may impart momentum to the surrounding fluid, potentially leading to the development of secondary flows. It is well established that secondary flow formation can significantly enhance the magnetophoretic transport of superparamagnetic nanoparticles~\cite{leong2015magnetophoresis,leong2020unified}. Moreover, these secondary flows may not only influence but also compete with field-induced particle cluster formation, potentially altering the overall separation dynamics. The strength of such secondary flows can be characterized by the dimensionless magnetic Grashof number $\mathrm{Gr}_m$. If $ \mathrm{Gr}_m < 1$, secondary flows are weak and have minimal impact on separation. Conversely, when $\mathrm{Gr}_m > 1$, secondary flows become significant, influencing the separation dynamics. The magnetic Grashof number $\mathrm{Gr}_m$ can be defined as~\cite{leong2020unified}:
\begin{equation}
\mathrm{Gr}_m = \frac{\rho_p \Delta \chi c_0 L^3}{\mathcal{M}_i (1 + \chi_{V,f}) \mu_0 \eta^2} |(\mathbf{B} \cdot \nabla) \mathbf{B}|_\mathrm{max},
\end{equation}  
Here $\mathcal{M}_i$, and $\chi_{V,f}$ are the molecular weight of the species in the solution, and the volumetric magnetic susceptibility of the fluid, respectively. Figure~(\ref{fig:grashof}) illustrates the magnetic Grashof number, $\mathrm{Gr}_m$, computed as a function of initial concentration and magnetic field gradients for the HGMS experimental parameters. There exist regions where $ \mathrm{Gr}_m > 1$, suggesting that secondary flows play a significant role in the separation process in this study. We hypothesize that metal ion separation in HGMS is governed by the interplay of two dominant mechanisms: (i) field-induced cluster formation and (ii) secondary flows. At low initial concentrations (e.g., $ c_0 = 1$ mM) where $\mathrm{Gr}_m < 1$, separation is primarily driven by field-induced cluster formation. However, as the initial concentration increases $\mathrm{Gr}_m $ may transition beyond unity, for which, secondary flows emerge as a significant factor enhancing separation. In this regime, separation efficiency likely benefits from the synergistic interaction of both mechanisms, potentially altering clustering dynamics. Thus, larger clusters are not necessarily required at higher concentrations; instead, secondary flows facilitate separation with smaller cluster sizes. In another example, at a fixed metal ion concentration (e.g., $c_0 = 100$ mM), increasing the applied magnetic field strength (or $|(\mathbf{B} \cdot \nabla) \mathbf{B}|_\mathrm{max}$) further amplifies secondary flow effects, reducing the need for large clusters to achieve efficient separation. The observed trends in $R_s$ values (Table~\ref{Table:ModelingParameters}) support this hypothesis. }
 

\subsection{Magnetophoretic Separation of Metal ions from Binary Mixtures}
In end-of-life LIBs, electronics, and many practical applications, metal ions with different magnetic properties often exist in mixtures, making it crucial to develop methods that enable their effective separation from each other within the mixture. Achieving selective separation of two ions based on magnetic properties could be particularly beneficial in such contexts. Therefore, in this section, we examined separation of paramagnetic MnCl$_2$ from diamagnetic  ZnCl$_2$ in a binary mixture of these ions. \par 

\begin{figure}[hbt!]
    \centering
    \includegraphics[width=0.32\linewidth]{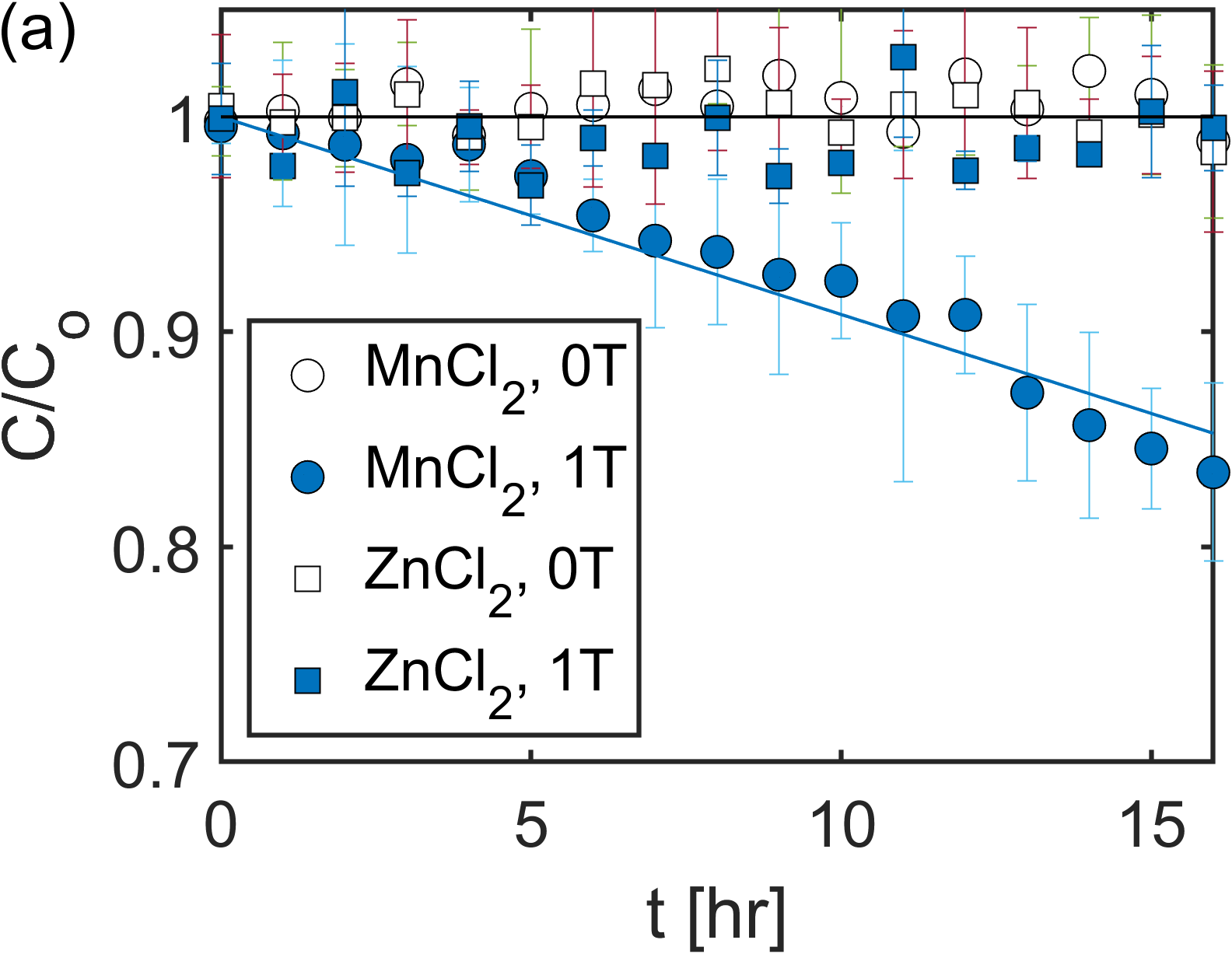}
    \includegraphics[width=0.32\linewidth]{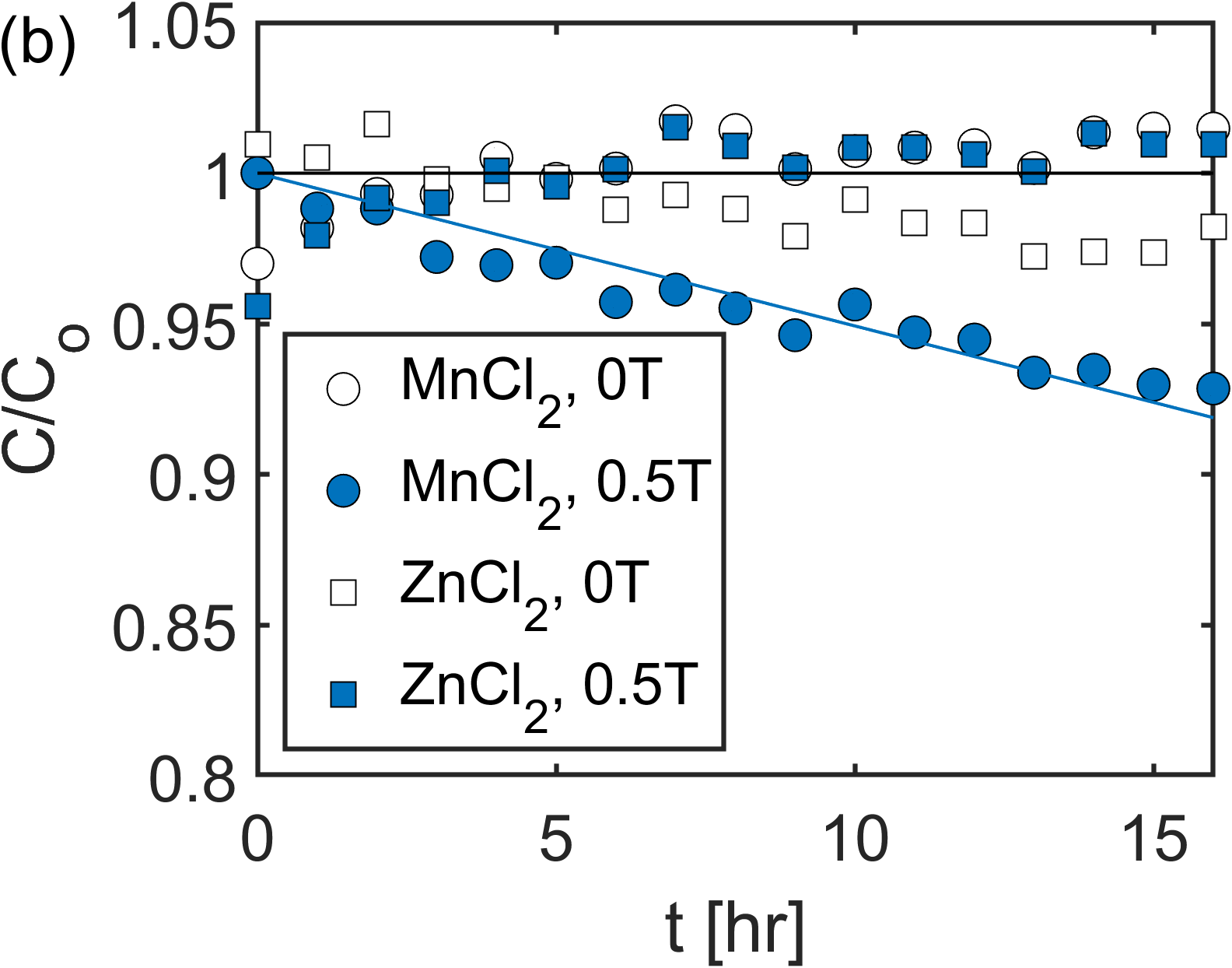}
    \includegraphics[width=0.32\linewidth]{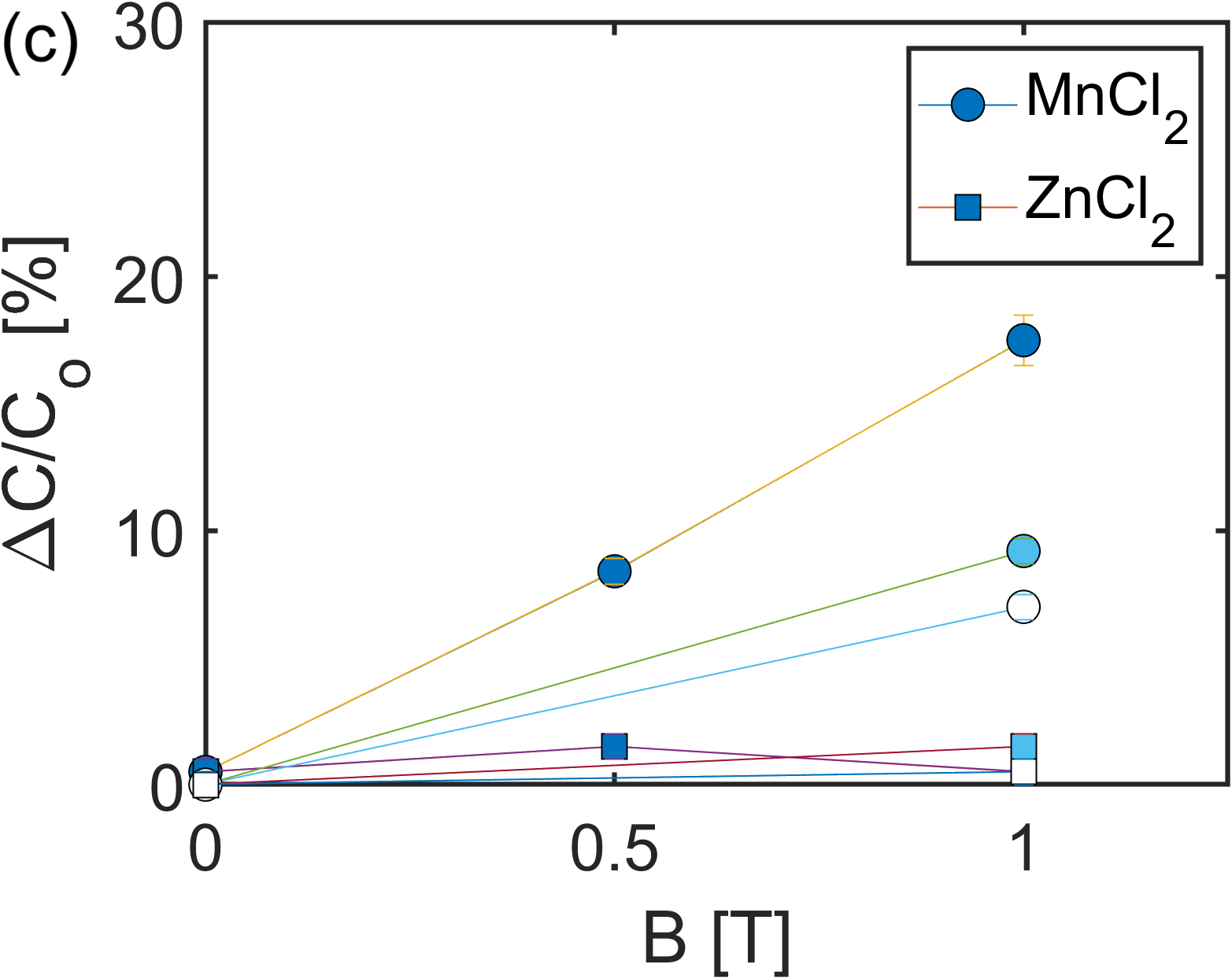}
    \caption{The normalized concentration of the MnCl$_2$ and ZnCl$_2$ metal ions as a function of time for binary mixtures containing both paramagnetic and diamagnetic ions at two magnetic field strengths: (a) $B$ = 1 T and (b) $B$ = 0.5 T. The overall concentration change (percentile) as a function of the applied magnetic field for both paramagnetic and diamagnetic metal ions is depicted for binary mixtures. In (c), the concentration of metal ions changes from 1mM, 10mM and 100mM for symbols from white to dark blue.}
    \label{fig:mixture}
\end{figure}

Fig.~\ref{fig:mixture}(a,b) show the normalized concentrations of MnCl$_2$ and ZnCl$_2$ in an equimolar mixture (100 mM) under varying magnetic fields. At a higher magnetic field strength of 1 T, MnCl$_2$ is effectively captured by the mesh, while the diamagnetic ions remain largely unaffected by the presence of the magnetic field or the paramagnetic metal ions. As a result, the diamagnetic ZnCl$_2$ ions exit the column at their initial concentration. Consistent with the single component solution experiments, increasing the magnetic field and the initial concentration of metal ions increases the capture rate of paramagnetic MnCl$_2$, as shown in Fig.~\ref{fig:mixture}(c). 
Perhaps equally important is the observation that the capture rate of the paramagnetic metal ion in a binary mixture is reduced compared to that measured in solutions containing individual paramagnetic or diamagnetic ions (c.f., Fig.~\ref{fig:mixture}(c) and Fig.~\ref{fig:singleelements}(f)). To model the separation process in the binary mixture, we assume that the effective magnetization of the mixture is a weighted average of the magnetization of the individual metal ions. This can be expressed as: $M_s = \sum x_i M_i$, where
$x_i$ represents the fraction of each metal ion in the mixture, and $M_i$ denotes the magnetization of each ion. For this case, 
$x_i = $0.5 for each component. Using this effective magnetization for the binary mixture, we have fitted the above model to the experimental data for mixtures under different magnetic field strengths. The resulting predictions are shown as continuous curves in Fig.~\ref{fig:mixture}(a,b). The critical radii values that produce the best match with mixture experiments turn out to be larger than those noted in Table~\ref{Table:ModelingParameters} ($R_s \approx$ 20.3 nm and 15 nm for $B$ = 0.5 T and $B$ = 1 T, respectively). These results underscore the impact of magnetic field strength on ion capture and highlight the complexity of separating metal ions in mixtures.\par

\section{Conclusion}
In summary, we report the first experimental demonstration of separating paramagnetic and diamagnetic metal ions, specifically MnCl$_2$ and ZnCl$_2$, from a solution mixture using high-gradient magnetic fields generated by a stainless steel mesh. This study highlights the potential for the use of magnetic fields to selectively capture metal ions on the basis of their magnetic properties. Our experiments were conducted on solutions containing either individual metal ions or binary aqueous mixtures. For solutions with single metal ions, the paramagnetic MnCl$_2$ is effectively attracted to the magnetic mesh wool, whereas the diamagnetic ZnCl$_2$ remains unaffected by the magnetic field gradients. We observed that increasing the initial concentration of the paramagnetic metal ion and the external magnetic field strength enhance the rate of magnetic capture.  \par 

In binary mixtures, paramagnetic metal ions are preferentially captured by the magnetic mesh wool, while diamagnetic ions were not retained. Additionally, the capture rate for paramagnetic metal ions increases as both the magnetic field strength and the initial ion concentration are raised. The capture rate of paramagnetic ions in binary mixtures is lower than that observed in solutions containing only single metal ions. We hypothesize that the presence of the diamagnetic ion in the mixture reduces the overall magnetization of the paramagnetic ion clusters because of the presence of Zn ions in the Mn ion clusters. This reduction in magnetization leads to a decreased attraction to the magnetic mesh wires, resulting in a lower capture rate for the paramagnetic ion. The inferred sizes of clusters are larger in binary mixtures of metal ions than in the individual ions of the same elements, which implies that the diamagnetic ion enters the ionic clusters of the paramagnetic ion, lowering the overall separation efficiency of HGMS.\par 

Furthermore, we modeled the magnetic separation of these metal ions using a multiphysics framework, which incorporate magnetic, viscous, and diffusive forces. Our modeling indicates that to achieve the observed levels of magnetic capture (10-50$\%$ depending on the conditions), the paramagnetic metal ions must aggregate into clusters significantly larger than their individual units approximately in the order of  $\approx \mathcal{O}(\SI{10})$ {nano-meter} under the influence of the magnetic field. \HM{Our analysis shows that in addition to field-induced cluster formation, secondary flows may be present near the wire surface and facilitate the separation of metal ions. Future studies should directly quantify the relative contributions of secondary flows and field-induced cluster formation in the magnetophoretic transport of metal ions near a wire surface.} We anticipate that these results will offer valuable insight into the recycling and recovery of critical metals from electronic devices. Additionally, they may open avenues for innovative strategies in various applications, such as the transport of ferromagnetic nanoparticles for drug delivery, protein separation, and water purification.\par

\section{Acknowledgement}
A portion of this work was performed at the National High Magnetic Field Laboratory, which is supported by the National Science Foundation Cooperative Agreement No. DMR-1644779 and the state of Florida. Additionally, HM gratefully acknowledges the support from the National Science Foundation through award CET 2343151. \HM{We greatly appreciate fruitful discussions with Peter Rassolov of the National High Magnetic Field Laboratory.}

\bibliography{Ref}

\bibliographystyle{unsrt}

\end{document}